\documentclass[aps,twocolumn,showpacs]{revtex4}

\usepackage{graphicx}
\usepackage{amsmath}
\usepackage{epsfig}
\usepackage{dcolumn}
\usepackage{bm}

\begin{document}

\title{Coherent Inverse Photoemission Spectrum for Gutzwiller Projected Superconductors} 

\author{Seiji Yunoki}

\affiliation{
Istituto Nazionale per la Fisica della Materia (INFM) and International 
School for Advanced Studies (SISSA), via Beirut 4, 34014 Trieste, Italy
}

\date{\today}

\begin{abstract}

Rigorous relations for Gutzwiller projected BCS states are derived. 
The obtained results do not depend on the details of model systems, but 
solely on the wave functions. Based on the derived relations, physical 
consequences are discussed for strongly correlated superconducting states 
such as high-$T_{\rm C}$ cuprate superconductors. 
\end{abstract}


\maketitle


Right after the discovery of high-$T_{\rm C}$ cuprate 
superconductors~\cite{bed}, Anderson has proposed a Gutzwiller projected 
BCS wave function --- a quantum many-body state incorporating strong on-site 
Coulomb repulsion --- to describe the superconducting state~\cite{pwa}. 
Since then there have been extensive studies in understanding the nature of 
this state and its variants~\cite{zhang,rvb}. In addition, several reports 
have shown that this projected BCS wave function is indeed a good 
variational ansatz state to describe the ground state of $t$-$J$ like 
models~\cite{gros,ogata,sandro,lee}, which are believed to capture the low 
energy physics of the cuprates~\cite{rice}. 
Although these projected BCS states were proposed more than 15 years ago, 
very recently they have acquired a revived 
interest~\cite{rand,anderson,ong,mohit}. This is probably because recent 
expensive numerical calculations based on the Gutzwiller projected variational 
ansatz clearly indicate that many aspects of the physics of high-$T_{\rm C}$ 
cuprate superconductors can be understood within this framework~\cite{rand}.

In this short communication some rigorous relations are derived for the 
Gutzwiller projected BCS states. 
It is shown that, as a consequence of the 
derived relations, the one-particle added excitation 
spectrum tends to be more coherent than the one-particle removed excitation 
spectrum does. It is further shown numerically that this trend is still 
observed approximately for more involved Gutzwiller projected BCS states. 
Possible experimental implications of the present results are also discussed.

Here our general system consists of a single orbital per unit cell on 
the two-dimensional (2D) square lattice with $L$ sites~\cite{lattice}.  
The creation and annihilation operators of spin 
$\sigma(=\uparrow,\downarrow)$ particle at site ${\bf i}$ are denoted by 
${\hat c}^\dag_{{\bf i}\sigma}$ and ${\hat c}_{{\bf i}\sigma}$, respectively. 
A Gutzwiller projected BCS state with $N$ particles is described by 
\begin{equation}
|\Psi^{(N)}_0\rangle = {\hat P}_N{\hat P}_G|{\rm BCS}\rangle,
\label{gs}
\end{equation}
where 
${\hat P}_N$ is the projection operator onto the fixed number $N$ of 
particles, 
${\hat P}_G=\prod_{\bf i}(1-{\hat n}_{{\bf i}\uparrow}{\hat n}_{{\bf i}\downarrow})$ 
is the Gutzwiller projection operator to restrict the Hilbert space with 
no double occupancy on each site, and 
${\hat n}_{{\bf i}\sigma}={\hat c}^\dag_{{\bf i}\sigma}
{\hat c}_{{\bf i}\sigma}$. 
$|{\rm BCS}\rangle=\prod_{{\bf k},\sigma} {\hat\gamma}_{{\bf k}\sigma}|0\rangle$ 
is the ground state of the BCS mean field Hamiltonian where
\begin{equation}
\left( 
 \begin{array}{c}
 {\hat\gamma}_{{\bf k}\uparrow}\\
 {\hat\gamma}^\dag_{-{\bf k}\downarrow}
 \end{array}
\right)
=
\left( 
 \begin{array}{cc}
 u_{\bf k}^* & -v_{\bf k}^*\\
 v_{\bf k} & u_{\bf k}
 \end{array}
\right)
\left( 
 \begin{array}{c}
 {\hat c}_{{\bf k}\uparrow}\\
 {\hat c}^\dag_{-{\bf k}\downarrow}
 \end{array}
\right)
\end{equation} 
are the standard Bogoliubov quasi-particle operators, 
${\hat c}_{{\bf k}\sigma}=\sum_{\bf i} 
{\rm e}^{-i{\bf k}\cdot{\bf i}}{\hat c}_{{\bf i}\sigma}/{\sqrt{L}}$, 
$|0\rangle$ is the vacuum of particles, and the singlet pairing is 
assumed~\cite{text}. The nature of this state has been extensively studied 
specially in the context of high-$T_{\rm C}$ 
cuprates~\cite{gros,ogata,sandro,lee,rand}

A one-particle added state with spin $\sigma$ and momentum ${\bf k}$ is 
similarly defined by using ${\hat\gamma}^\dag_{{\bf k}\sigma}$:  
\begin{equation}
|\Psi^{(N+1)}_{{\bf k}\sigma}\rangle = {\hat P}_{N+1}{\hat P}_G
{\hat\gamma}^\dag_{{\bf k}\sigma}|{\rm BCS}\rangle.
\label{est}
\end{equation}
This state was first proposed by Zhang, {\it et al}~~\cite{zhang}, followed 
by several others~\cite{rand,yunoki,ong}. 
Hereafter the normalized wave functions for the $N$- and 
$(N+1)$-particle states are denoted by $|\psi^{(N)}_0\rangle$ and 
$|\psi^{(N+1)}_{{\bf k}\sigma}\rangle$, respectively.

First it is useful to show that the following operator relation between 
${\hat c}_{{\bf k}\sigma}$ and ${\hat c}^\dag_{{\bf k}\sigma}$ holds exactly: 
\begin{equation}
{\hat P}_G{\hat c}_{{\bf k}\sigma}{\hat c}^\dag_{{\bf k}\sigma}{\hat P}_G
={\hat c}_{{\bf k}\sigma}{\hat P}_G{\hat c}^\dag_{{\bf k}\sigma}
+ {\frac{1}{L}}{\hat N}_{\bar{\sigma}}{\hat P}_G. 
\label{eq1}
\end{equation} 
Here 
${\hat N}_{\sigma}=\sum_{\bf i}{\hat c}^\dag_{{\bf i}\sigma}{\hat c}_{{\bf i}\sigma}$ 
and ${\bar\sigma}$ stands for the opposite spin of $\sigma$. 
This is easily 
proved by using ${\hat P}_G{\hat c}^\dag_{{\bf i}\sigma}{\hat P}_G =
 {\hat P}_G{\hat c}^\dag_{{\bf i}\sigma}$.

Using the above equation (\ref{eq1}), it is readily shown that the momentum 
distribution function 
$n_\sigma({\bf k})=\langle\psi^{(N)}_0|{\hat c}^\dag_{{\bf k}\sigma}
{\hat c}_{{\bf k}\sigma}|\psi^{(N)}_0\rangle$ calculated for the state 
$|\Psi^{(N)}_0\rangle$ is related to 
the state $|\Psi^{(N+1)}_{{\bf k}\sigma}\rangle$ through 
\begin{equation}
n_\sigma({\bf k})=1-{\frac{N_{\bar\sigma}}{L}}-|u_{\bf k}|^2 
{\frac{\langle\Psi^{(N+1)}_{{\bf k}\sigma}|\Psi^{(N+1)}_{{\bf k}\sigma}\rangle}
{\langle\Psi^{(N)}_0|\Psi^{(N)}_0\rangle}}, 
\label{nk}
\end{equation}
where ${\hat N}_{\sigma}{\hat P}_G={\hat P}_G{\hat N}_{\sigma}$ is used.

The quasi-particle weight for the one-particle added excitation is defined by 
\begin{equation}
Z_{{\bf k}\sigma}^{(+)}=\left|\langle\psi^{(N+1)}_{{\bf k}\sigma}|
{\hat c}^\dag_{{\bf k}\sigma}|\psi^{(N)}_0\rangle\right|^2.
\label{z}
\end{equation} 
Now we shall show that there exists a simple and exact relation between 
$Z_{{\bf k}\sigma}^{(+)}$ and $n_\sigma({\bf k})$. To this end,  
it is important to notice that since 

\begin{equation}
{\langle\Psi^{(N+1)}_{{\bf k}\sigma}|
{\hat c}^\dag_{{\bf k}\sigma}|\Psi^{(N)}_0\rangle}
=u_{\bf k}^*{\langle\Psi^{(N+1)}_{{\bf k}\sigma}|\Psi^{(N+1)}_{{\bf k}\sigma}\rangle}
\end{equation}
$Z_{{\bf k}\sigma}^{(+)}$ is simplified as 
\begin{equation}
Z_{{\bf k}\sigma}^{(+)}=|u_{\bf k}^2|{\frac{\langle\Psi^{(N+1)}_{{\bf k}\sigma}|\Psi^{(N+1)}_{{\bf k}\sigma}\rangle}{\langle\Psi^{(N)}_0|\Psi^{(N)}_0\rangle}}. 
\label{zk+}
\end{equation}
From Eqs.~(\ref{nk}) and (\ref{zk+}), we finally arrive at the desired 
relation, 
\begin{equation}
n_\sigma({\bf k})+Z_{{\bf k}\sigma}^{(+)}=1-{\frac{N_{\bar\sigma}}{L}}. 
\label{exact}
\end{equation}
It should be emphasized that to derive the above equation we have not 
made either any approximations or any assumptions except for the form of 
the wave functions for the $N$- and $(N+1)$-particle states given by 
Eqs.~(\ref{gs}) and (\ref{est}), respectively.

\begin{figure}[hbt]
\includegraphics[clip=true,width=8.cm,angle=-0]{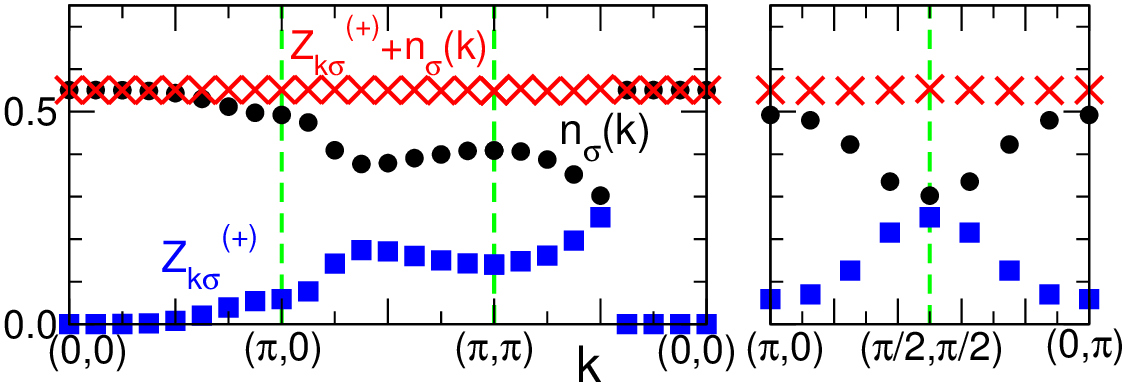}
\begin{center}
\caption{
$n_\sigma({\bf k})$ (circles) and $Z_{{\bf k}\sigma}^{(+)}$ (squares) 
calculated using a Monte Carlo technique for $L=16\times16$ and 
$N_{\uparrow}=N_{\downarrow}=115$~\cite{note}. 
The sum of the two quantities 
($n_\sigma({\bf k})+Z_{{\bf k}\sigma}^{(+)}$) is also 
plotted by crosses, which are $1-N_{\bar\sigma}/L=0.551$ within the 
statistical error bars (smaller than the size of the symbols). 
}
\label{sum}
\end{center}
\end{figure}

The equation (\ref{exact}) is also simply verified numerically on small 
clusters using a Monte Carlo technique. 
Typical results are presented in Fig.~\ref{sum}. As seen in Fig.~\ref{sum} 
and for all other cases studied, Eq.~(\ref{exact}) is satisfied within the 
statistical errors.

In order to discuss a physical consequence of Eq.~(\ref{exact}) on the 
one-particle excitation spectrum, let us first derive a simple same rule. 
The one-particle excitation spectra for removing one particle 
$[A_\sigma^{\rm PES}({\bf k},\omega)]$ and for adding one particle 
$[A_\sigma^{\rm IPES}({\bf k},\omega)]$ are defined respectively by 
\begin{eqnarray}
A_\sigma^{\rm PES}({\bf k},\omega) &=& -{\frac{1}{\pi}}
{\rm Im} \left\langle {\hat c}_{{\bf k}\sigma}^\dag 
  {\frac{1}{\omega+({\hat H}-E_0)+i0^+}}{\hat c}_{{\bf k}\sigma}
\right \rangle \nonumber\\
A_\sigma^{\rm IPES}({\bf k},\omega) &=& -{\frac{1}{\pi}}
{\rm Im} \left\langle {\hat c}_{{\bf k}\sigma} 
  {\frac{1}{\omega-({\hat H}-E_0)+i0^+}}{\hat c}_{{\bf k}\sigma}^\dag
\right \rangle \nonumber,
\end{eqnarray}
where $\langle{\hat O}\rangle$ is the expectation value of ${\hat O}$ 
for the exact ground state of the system described by Hamiltonian ${\hat H}$ 
with its eigenvalue $E_0$. 
It is generally proved that for any model systems defined within the 
restricted Hilbert space with no double occupancy on any sites by particles, 
such as the $t$-$J$ model, the 0-th moment of the spectral function satisfies 
the following sum rules: 
\begin{eqnarray}
\int_{-\infty}^\infty A_{\sigma}^{\rm PES}({\bf k},\omega)d\omega&=
&\langle{\hat c}_{{\bf k}\sigma}^\dag{\hat c}_{{\bf k}\sigma}\rangle \label{pes}\\
\int_{-\infty}^\infty A_{\sigma}^{\rm total}({\bf k},\omega)d\omega&=&1-N_{\bar\sigma}/L, \label{total}
\end{eqnarray}
where $A_{\sigma}^{\rm total}({\bf k},\omega)=A_\sigma^{\rm PES}({\bf k},\omega)+A_\sigma^{\rm IPES}({\bf k},\omega)$. The latter equation is easily proved 
by using Eq.~(\ref{eq1}). Eq.~(\ref{pes}) is a rather standard sum rule, 
while Eq.~(\ref{total}) is due to the reduction of the Hilbert space by 
${\hat P}_G$ and it is indeed satisfied for, {\it e.g.}, the $t$-$J$ 
model~\cite{horsch}.

Let us now discuss what physical consequences would be expected. 
First we assume that there exists a system for which the ground 
state and the low-lying excited states are approximately described by the 
wave functions, Eqs.~(\ref{gs}) and (\ref{est}), introduced above. Then it 
follows immediately from Eqs.~(\ref{exact})--(\ref{total}) that 
\begin{equation}
\int_{-\infty}^\infty A_{\sigma}^{\rm IPES}({\bf k},\omega)d\omega=
Z_{{\bf k}\sigma}^{(+)}.
\label{sumrule}
\end{equation}
This relation implies that the one-particle added excitation spectrum 
is all coherent since only one state contributes to 
$A_{\sigma}^{\rm IPES}({\bf k},\omega)$. 
It should be noted here that while several studies have recently reached the 
similar conclusions~\cite{ong,mohit}, 
the argument presented here is more rigorous and transparent.

\begin{figure}[hbt]
\includegraphics[clip=true,width=8.cm,angle=-0]{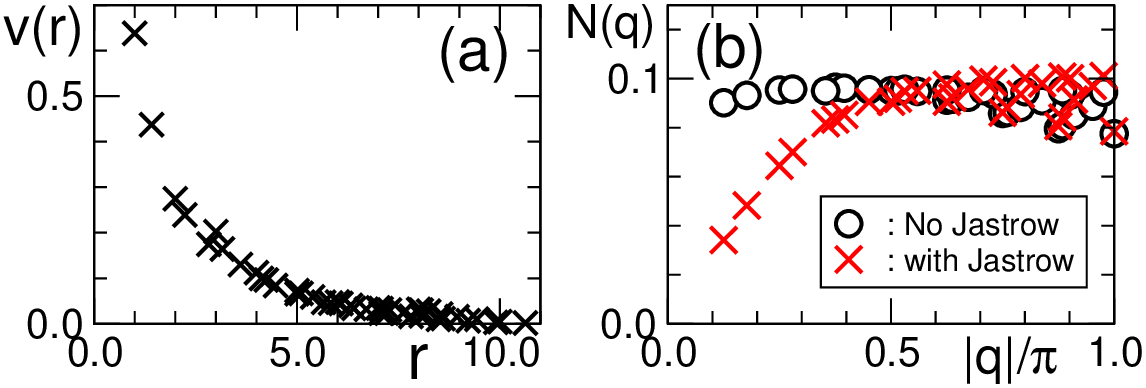}
\begin{center}
\caption{
(a) Charge Jastrow factor $v(r)=v_{\bf ij}$ as a function of distances 
$r=|{\bf i}-{\bf j}|$. These quantities are optimized in such a way that 
the variational energy of $|\Phi^{(N)}_0\rangle$ is minimized for the 2D 
$t$-$t'$-$J$ model with $J/t=0.3$, $t'/t=-0.2$, and 
$N_{\uparrow}=N_{\downarrow}=115$ on $L=16\times16$~\cite{yunoki,note}. 
(b) Charge structure factor $N({\bf q})$ calculated using 
$|\Psi^{(N)}_0\rangle$ 
(circles) and $|\Phi^{(N)}_0\rangle$ (crosses) for the 2D $t$-$t'$-$J$ 
model with the same model parameters as in (a). The variational parameters 
are optimized for both states. 
The statistical error bars are smaller than the size of the symbols. 
}
\label{jastrow}
\end{center}
\end{figure}

Next we shall discuss to what extend Eq.~(\ref{exact}) holds and therefore 
Eq.~(\ref{sumrule}) remains approximately true for more involved wave 
functions. 
A natural and important extension of the simplest Gutzwiller projected BCS 
states described by Eqs.~(\ref{gs}) and (\ref{est}) can be achieved by 
including charge Jastrow factors ${\hat J}_C$, 
{\it i.e.}, 
\begin{eqnarray}
|\Phi^{(N)}_0\rangle &=& {\hat P}_N{\hat P}_G{\hat J}_C|{\rm BCS}\rangle,
\label{gsj}\\
|\Phi^{(N+1)}_{{\bf k}\sigma}\rangle &=& {\hat P}_{N+1}{\hat P}_G
{\hat J}_C{\hat\gamma}^\dag_{{\bf k}\sigma}|{\rm BCS}\rangle, 
\label{estj}
\end{eqnarray} 
for the ground state and the one-particle excited states, respectively. 
Here 
\begin{equation}
{\hat J}_C = \exp\left(-\sum_{{\bf i},{\bf j}}v_{\bf ij}{\hat n}_{\bf i}
{\hat n}_{\bf j}\right),
\end{equation}
${\hat n}_{\bf i}={\hat n}_{{\bf i}\uparrow}+{\hat n}_{{\bf i}\downarrow}$, 
and the sum runs over all the independent pairs of sites {\bf i} and {\bf j}. 
The importance of ${\hat J}_C$ has been already reported for various lattice 
models~\cite{sandro,capello}. A typical example of $v_{\bf ij}$ is presented 
in Fig.~\ref{jastrow} (a) where all the independent $v_{\bf ij}$ are 
optimized for the 2D $t$-$t'$-$J$ model with $J/t=0.3$, $t'/t=-0.2$, and 
$N_{\uparrow}=N_{\downarrow}=115$ on $L=16\times16$~\cite{yunoki}.
Certainly the inclusion of ${\hat J}_C$ improves, {\it e.g.}, the 
variational energy. 
Besides such quantitative changes, ${\hat J}_C$ can also make a qualitative 
difference. 
One of these examples is shown in Fig.~\ref{jastrow} (b), where 
the charge structure factor 
$N({\bf q})=\sum_{\bf l}\exp(-i{\bf q}\cdot{\bf l})\langle{\hat n}_{\bf i}
{\hat n}_{{\bf i}+{\bf l}} \rangle$ for small wave numbers are calculated 
using the wave functions with and without ${\hat J}_C$. As seen in 
Fig.~\ref{jastrow} (b), 
$N({\bf q})\to0$ as $|{\bf q}|\to0$ for $|\Phi^{(N)}_0\rangle$ as expected, 
whereas $N({\bf q})\to$ finite as $|{\bf q}|\to0$ for 
$|\Psi^{(N)}_0\rangle$~\cite{note2}. 
This is because $|{\rm BCS}\rangle$ does not conserve the number of 
particles which is instead a conserved quantity for the $t$-$t'$-$J$ 
model~\cite{note6,note3}.

\begin{figure}[hbt]
\includegraphics[clip=true,width=8.cm,angle=-0]{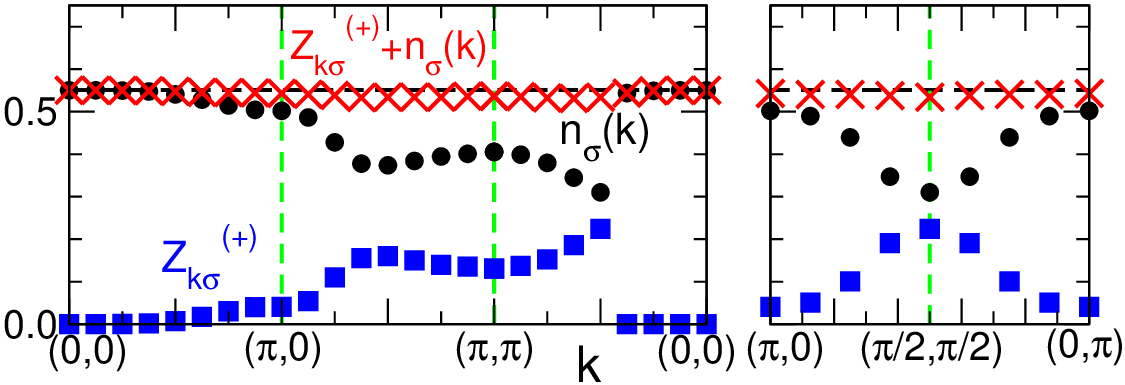}
\begin{center}
\caption{
$n_\sigma({\bf k})$ (circles) and $Z_{{\bf k}\sigma}^{(+)}$ (squares) 
calculated using the wave functions $|\Phi^{(N)}_0\rangle$ and 
$|\Phi^{(N+1)}_{{\bf k}\sigma}\rangle$ with ${\hat J}_C$. The model 
parameters used are the same as in Fig.~\ref{jastrow}. 
The sum of the two quantities 
($n_\sigma({\bf k})+Z_{{\bf k}\sigma}^{(+)}$) and $1-N_{\bar\sigma}/L=0.551$ 
are also plotted by crosses and thick dashed lines, respectively. 
The statistical error bars are smaller than the size of the symbols. 
}
\label{sum2}
\end{center}
\end{figure}

It is now interesting to examine if the exact relation Eq.~(\ref{exact}) 
proved for the states $|\Psi^{(N)}_0\rangle$ and 
$|\Psi^{(N+1)}_{{\bf k}\sigma}\rangle$, and thus Eq.~(\ref{sumrule}),  
can still hold for these more involved wave functions 
$|\Phi^{(N)}_0\rangle$ and $|\Phi^{(N+1)}_{{\bf k}\sigma}\rangle$. 
The numerical results on small clusters are presented in Fig~\ref{sum2} 
for the 2D $t$-$t'$-$J$ 
model with the same model parameters as in Fig.~\ref{jastrow}. 
As seen in Fig~\ref{sum2}, surprisingly Eq.~(\ref{exact}) remains 
satisfied, at least approximately. Thus it can be still argued that, 
because of the sum rule for the one-particle excitation spectrum 
[Eqs.~(\ref{pes}) and (\ref{total})], Eq.~(\ref{sumrule}) is 
approximately satisfied, and therefore most of the one-particle added 
excitation spectrum consists of a coherent part. 
Further numerical calculations have been carried out for the 2D $t$-$t'$-$J$ 
model with different model parameters, and 
it was found that Eq.~(\ref{exact}) is still satisfied 
within 10--15$\%$~\cite{note4}.

So far we have only considered the one-particle added excitations. 
Let us briefly discuss the one-particle removed excitations. 
A one-particle removed state is analogously constructed by 
\begin{equation}
|\Psi^{(N-1)}_{{\bf k}\sigma}\rangle = {\hat P}_{N-1}{\hat P}_G
{\hat\gamma}^\dag_{{\bf k}\sigma}|{\rm BCS}\rangle.
\end{equation} 
Although $|\Psi^{(N-1)}_{{\bf k}\sigma}\rangle$ and 
$|\Psi^{(N+1)}_{{\bf k}\sigma}\rangle$ are of very like form, the similar 
conclusion about the coherence of the one-particle excitations can not 
be drawn for the one-particle removed excitations. 
A simple reason for this is the following~\cite{ong,mohit}: 
for the one-particle added excitations, 
$$
{\hat P}_G{\hat c}_{{\bf k}\sigma}^\dag|\Psi^{(N)}_0\rangle
={\hat P}_{N+1}{\hat P}_G{\hat c}_{{\bf k}\sigma}^\dag|{\rm BCS}\rangle 
\propto |\Psi^{(N+1)}_{{\bf k}\sigma}\rangle,
$$ 
{\it i.e.}, ${\hat P}_G{\hat c}_{{\bf k}\sigma}^\dag|\Psi^{(N)}_0\rangle$ 
consists of only one state, while for the one-particle removed excitations 
${\hat P}_G{\hat c}_{{\bf k}\sigma}|\Psi^{(N)}_0\rangle
={\hat c}_{{\bf k}\sigma}|\Psi^{(N)}_0\rangle$, which is not described by 
$|\Psi^{(N-1)}_{-{\bf k}{\bar{\sigma}}}\rangle$ alone. It is also 
checked numerically on small clusters that the quasi-particle weight for 
the one-particle removed excitation, 
$Z_{{\bf k}\sigma}^{(-)}=\left|\langle\psi^{(N-1)}_{-{\bf k}{\bar \sigma}}|
{\hat c}_{{\bf k} \sigma}|\psi^{(N)}_0\rangle\right|^2$, is 
substantially different from $n_{\sigma}({\bf k})$.

Finally we shall discuss experimental implications of the present results. 
A most relevant experiment is angle-resolved {\it inverse} photoemission 
spectroscopy on the {\it superconducting state} for the {\it hole-doped} 
cuprates. 
If we assume that the Gutzwiller projected BCS states discussed here are 
faithful description for the superconducting state in the cuprates, 
it is expected that the inverse photoemission spectroscopy spectrum has more 
coherent characteristics than the direct photoemission spectroscopy spectrum 
does~\cite{note5}. The similar trend is also expected in the 
{\it superconducting state} for the {\it electron-doped} cuprates except now 
that the {\it direct} photoemission spectroscopy spectrum has more coherent 
characteristics. This is because the $t$-$J$ like models can also describe 
the electron-doped cuprates only after the particle-hole transformation: 
${\hat c}_{{\bf k}\sigma}\to{\hat h}_{{{\bf k}+{\bf Q}}\sigma}^\dag$ 
[${\bf Q}=(\pi,\pi)$]~\cite{tohyama}, and the same argument presented here is 
still true for ${\hat h}_{{\bf k}\sigma}^\dag$.

To summarize, we have derived some rigorous relations for the Gutzwiller 
projected BCS states. Using a sum rule for the one-particle excitation 
spectrum, it was shown that the one-particle added excitation spectrum 
tends to be more coherent than the one-particle removed excitation does. 
Possible experimental implications were also discussed. 
Finally, it should be noted that all the results presented here are based 
on the Gutzwiller projected BCS states studied, and a question of 
whether these states can represent the exact eigenstates of some particular 
model Hamiltonian is beyond the scope of the present study.


The author is grateful to S. Sorella, R. Hlubina, and E. Dagotto for 
stimulating discussions. This work was supported in part by 
INFM through contract n. OA04007678.

%
%
%



\begin{thebibliography}{99}
\bibitem{lattice} Even though the 2D square lattice is considered in this 
paper, the results presented here for the projected BCS states without charge 
Jastrow factor ${\hat J}_C$ can be also applied on any spatial 
dimensions with any lattice geometry.
\bibitem{note} Following the standard BCS theory with $d$-wave 
pairing~\cite{text}, 
$|u_{\bf k}|^2=\left[1+\xi_{\bf k}/E_{\bf k}\right]/2$ and 
$|v_{\bf k}|^2=\left[1-\xi_{\bf k}/E_{\bf k}\right]/2$ are parameterized 
by $E_{\bf k}={\sqrt{\xi_{\bf k}^2+|\Delta_{\bf k}|^2}}$, 
$\xi_{\bf k}=-2t(\cos k_x+\cos k_y)-4t^\prime\cos 2k_x \cos 2k_y-\mu$, and 
$\Delta_{\bf k}=\Delta(\cos k_x-\cos k_y)$. For the calculations presented 
in Fig.~\ref{sum}, $\Delta/t=0.448$, $t^\prime/t=-0.302$, and 
$\mu/t=-0.675$ are chosen, 
which are optimized variational parameters for the 2D $t$-$t'$-$J$ model with 
$J/t=0.3$ and $t'/t=-0.2$~\cite{yunoki}.
\end{thebibliography}
\end{document}